\documentclass[a4paper,aps,twocolumn,nofootinbib]{revtex4}
\RequirePackage[colorlinks,hyperindex]{hyperref}
\RequirePackage[english]{babel}
\RequirePackage[latin1]{inputenc}
\RequirePackage[T1]{fontenc}
\RequirePackage{mathrsfs}
\RequirePackage{amsmath}
\RequirePackage{amssymb}
\RequirePackage{amsbsy}
\RequirePackage{color}
\RequirePackage{bm}
\hypersetup{colorlinks=true,breaklinks=true,urlcolor=blue,linkcolor=red}
\begin{document}
%%%%%%%%%%%%%%%%%%%%%%%%%%%%%%%%%%%%%%%%%%%%%%%%%%%%%%%%%%%%%%%%%%%%%%%%%%%%%%%%%%%%%%%%%%%%%%%%%%%
\title{\bf{Re-normalizable Chern-Simons Extension of Propagating Torsion Theory}}
\author{Luca Fabbri}
\affiliation{DIME, Sez. Metodi e Modelli Matematici, Universit\`{a} di Genova,\\
Via all'Opera Pia 15, 16145 Genova, ITALY}
\date{\today}
%%%%%%%%%%%%%%%%%%%%%%%%%%%%%%%%%%%%%%%%%%%%%%%%%%%%%%%%%%%%%%%%%%%%%%%%%%%%%%%%%%%%%%%%%%%%%%%%%%%
\begin{abstract}
We write the most general parity-even re-normalizable Chern-Simons term for massive axial-vector propagating torsion fields. After obtaining the most comprehensive action, we perform the causal structure analysis to see what self-interaction term must be suppressed. In view of such a restriction for the Lagrangian, we will obtain the field equations, investigating some of their properties.
\end{abstract}
%%%%%%%%%%%%%%%%%%%%%%%%%%%%%%%%%%%%%%%%%%%%%%%%%%%%%%%%%%%%%%%%%%%%%%%%%%%%%%%%%%%%%%%%%%%%%%%%%%%
\maketitle
%%%%%%%%%%%%%%%%%%%%%%%%%%%%%%%%%%%%%%%%%%%%%%%%%%%%%%%%%%%%%%%%%%%%%%%%%%%%%%%%%%%%%%%%%%%%%%%%%%%
\section{Introduction}
In the Cartan extension of Riemann geometry the most general connection which is compatible with the metric is not constrained to be symmetric in the two lower indices, with the result that the torsion tensor can be defined in concomitance to the curvature tensor of the space-time. 

And the Sciama-Kibble theory of torsion is the natural completion of Einstein gravity in which torsion is sourced by the spin in the same way in which curvature is sourced by the energy density of the matter field distributions.

For a general review on torsion-gravity we refer to the excellent reports we listed as references \cite{Hehl:1976kj}, \cite{Shapiro:2001rz} and \cite{Hammond:2002rm,Arcos:2005ec}.

There are two features in terms of which torsion gravity becomes particularly interesting, the first of which being the fact that there are general arguments indicating that torsion should be taken completely antisymmetric \cite{Laemmerzahl:1993zn,Audretsch:1988tu,Fabbri:2006xq,Fabbri:2009se}.

Because torsion couples to the spin, such a restriction implies that we can only work with completely antisymmetric spin, and so the $1/2$-spin spinor field solely \cite{Fabbri:2008rq,Fabbri:2009yc}.

This is a fortunate coincidence, since of all matter field distributions, the $1/2$-spin spinor field, that is the Dirac field, seems to be the only one that is found in nature.

The second feature that makes torsion gravity interesting is that, when torsion is taken as massive propagating field, it becomes clear that the torsion is not necessarily confined at the Planck scale, dissolving a misconception that was thought to afflict such a theory for a long time.

Additionally, torsion couples to the spin according to a different coupling constant for different spinor fields, and as a matter of fact one can be allowed to think at torsion for spinors as a sort of spin-dependent Higgs field \cite{Fabbri:2014naa}.

The fact that torsion can propagate also has the consequence that even if taken in the effective approximation the torsion-spin coupling constant has fixed negative sign, making up for the interpretation of torsion as an effective interaction of attractive type for the spinor with itself, or more specifically between its two irreducible chiral parts.

Combining the fact that in Einsteinian gravitation the curvature of the space-time couples to the energy, which differently from the mass can also account for a potential term and therefore be negative, with the fact that such a torsion-spin coupling is attractive, which means that we get indeed potential contributions that are negative and can become dominant at large densities, gives rise to the result that gravitationally-induced singularity formation may be prevented by this torsion-spin interaction \cite{Fabbri:2017rjf,Fabbri:2017xch}.

This solves another problem that was thought to affect all gravitational theories, and even in conditions in which gravity were weak enough to be neglected, a torsion-spin coupling still provides some insights regarding the exclusion principle and matter/antimatter duality \cite{Fabbri:2010rw,Fabbri:2012ag}.

The most general theory of gravity with torsion in presence of spinor fields compatible with all requirements on causal propagation has been developed in \cite{Fabbri:2014dxa}.

This theory, however, does not take into account some terms of topological origin. The principal reason for this omission is that these terms would have to be written in terms of multiplicative pseudo-scalar factors that would spoil the re-normalizability of the theory as a whole.

This occurrence is due to the fact that a pseudo-scalar field has unitary mass dimension since it is introduced as a field obeying the second-order derivative Klein-Gordon equation. Nevertheless, there would be none of the above problems if it were possible to find one pseudo-scalar field undergoing first-order derivative Dirac-like equations.

As uncommon as this might look, recent developments in spinor field theory have highlighted the existence of a pseudo-scalar field of this sort. Because it will have zero mass dimension it is suitable for writing re-normalizable topological terms that would not spoil causal propagation of torsion while allowing an extension of such a theory.

In this paper we are going to explain how to introduce this possibility. In a later part of the paper we will study some general property of such an extended theory.
%%%%%%%%%%%%%%%%%%%%%%%%%%%%%%%%%%%%%%%%%%%%%%%%%%%%%%%%%%%%%%%%%%%%%%%%%%%%%%%%%%%%%%%%%%%%%%%%%%%
%%%%%%%%%%%%%%%%%%%%%%%%%%%%%%%%%%%%%%%%%%%%%%%%%%%%%%%%%%%%%%%%%%%%%%%%%%%%%%%%%%%%%%%%%%%%%%%%%%%
\section{Spinor Fields}
We begin by recalling and fixing notations and conventions. Clifford matrices $\boldsymbol{\gamma}^{a}$ are defined as $\left\{\boldsymbol{\gamma}_{a}\!,\!\boldsymbol{\gamma}_{b}\right\}\!=\!2\eta_{ab}\mathbb{I}$ where $\eta_{ab}$ is the Minkowski matrix. Hence $\left[\boldsymbol{\gamma}_{a}\!,\!\boldsymbol{\gamma}_{b}\right]\!=\!4\boldsymbol{\sigma}_{ab}$ defines the generators of the complex Lorentz algebra and we have that the relationship $2i\boldsymbol{\sigma}_{ab}\!=\!\varepsilon_{abcd}\boldsymbol{\pi}\boldsymbol{\sigma}^{cd}$ implicitly defines the $\boldsymbol{\pi}$ matrix (this matrix is usually denoted as a gamma matrix with an index five, but in space-time this index has no meaning, and so we employ a notation with no index). By exponentiating the generators $\boldsymbol{\sigma}_{ab}$ it is possible to find the local complex Lorentz group $\boldsymbol{S}$ and a spinor field $\psi$ is defined as what transforms according to $\psi\!\rightarrow\!\boldsymbol{S}\psi$ in general. With the Clifford matrices we can also build a procedure that will convert a spinor $\psi$ in its adjoint spinor $\overline{\psi}\!=\! \psi^{\dagger}\boldsymbol{\gamma}^{0}$ again in general. With a Clifford basis, and the pair of adjoint spinors, we can construct a set of bi-linear spinor quantities according to
\begin{eqnarray}
&\Sigma^{ab}\!=\!2\overline{\psi}\boldsymbol{\sigma}^{ab}\boldsymbol{\pi}\psi\\
&M^{ab}\!=\!2i\overline{\psi}\boldsymbol{\sigma}^{ab}\psi
\end{eqnarray}
\begin{eqnarray}
&S^{a}\!=\!\overline{\psi}\boldsymbol{\gamma}^{a}\boldsymbol{\pi}\psi\\
&U^{a}\!=\!\overline{\psi}\boldsymbol{\gamma}^{a}\psi
\end{eqnarray}
\begin{eqnarray}
&\Theta\!=\!i\overline{\psi}\boldsymbol{\pi}\psi\\
&\Phi\!=\!\overline{\psi}\psi
\end{eqnarray}
which are normalized as to be all real and they are such that $U_{a}U^{a}\!=\!-S_{a}S^{a}\!=\!|\Theta|^{2}\!+\!|\Phi|^{2}$ and $U_{a}S^{a}\!=\!0$ hold.

These bi-linears can be used to perform a classification of spinor fields known as Lounesto classification \cite{L, Cavalcanti:2014wia}: singular spinors are those for which $\Theta\!=\!\Phi\!\equiv\!0$ and such a class contains the flag-dipole, flagpole and dipole spinors, discussed in \cite{HoffdaSilva:2017waf, daSilva:2012wp, daRocha:2008we, Ablamowicz:2014rpa, Rodrigues:2005yz, daRocha:2013qhu, HoffdaSilva:2009is} and references therein; nevertheless, our interest will be on regular spinors defined when not both $\Theta$ and $\Phi$ are equal to zero identically. In such a case it is possible to demonstrate that spinors can always be written, in chiral representation, according to
\begin{eqnarray}
&\!\psi\!=\!\phi e^{-\frac{i}{2}\beta\boldsymbol{\pi}}
\boldsymbol{S}\left(\!\begin{tabular}{c}
$1$\\
$0$\\
$1$\\
$0$
\end{tabular}\!\right)
\label{spinor}
\end{eqnarray}
for some complex Lorentz transformation $\boldsymbol{S}$ with $\phi$ and $\beta$ called module and Yvon-Takabayashi angle, and where the spinor is said to be in polar form \cite{Fabbri:2016msm}. In taking the polar form into the bi-linear spinor quantities we get
\begin{eqnarray}
&\Sigma^{ab}\!=\!2\phi^{2}(\cos{\beta}u^{[a}s^{b]}\!-\!\sin{\beta}u_{j}s_{k}\varepsilon^{jkab})\\
&M^{ab}\!=\!2\phi^{2}(\cos{\beta}u_{j}s_{k}\varepsilon^{jkab}\!+\!\sin{\beta}u^{[a}s^{b]})
\end{eqnarray}
showing that they are not independent as they can always be written with the vector bi-linear spinor quantities
\begin{eqnarray}
&S^{a}\!=\!2\phi^{2}s^{a}\\
&U^{a}\!=\!2\phi^{2}u^{a}
\end{eqnarray}
and the scalar bi-linear spinor quantities
\begin{eqnarray}
&\Theta\!=\!2\phi^{2}\sin{\beta}\\
&\Phi\!=\!2\phi^{2}\cos{\beta}
\end{eqnarray}
such that $u_{a}u^{a}\!=\!-s_{a}s^{a}\!=\!1$ and $u_{a}s^{a}\!=\!0$ and which show that module and Yvon-Takabayashi angle are the only $2$ true degrees of freedom. The $8$ real components of spinors are rearranged into the special configuration in which the $2$ real scalar degrees of freedom, the module and the YT angle, are isolated from the $6$ components that can always be transferred away, the spin and velocity. We notice also that the YT angle is a zero-dimension pseudo-scalar, and therefore the module inherits the full $3/2$-dimension that characterizes the spinor field. This is the most important remark for the following of the paper, as we shall see. 

For the background, we have that with the metric we can define the symmetric connection $\Lambda^{\sigma}_{\alpha\nu}$ and with it we define the spin connection $\Omega^{a}_{\phantom{a}b\pi}\!=\!\xi^{\nu}_{b}\xi^{a}_{\sigma}(\Lambda^{\sigma}_{\nu\pi}\!-\!\xi^{\sigma}_{i}\partial_{\pi}\xi_{\nu}^{i})$ so that with the gauge potential $qA_{\mu}$ we have
\begin{eqnarray}
&\boldsymbol{\Omega}_{\mu}
=\frac{1}{2}\Omega^{ab}_{\phantom{ab}\mu}\boldsymbol{\sigma}_{ab}
\!+\!iqA_{\mu}\boldsymbol{\mathbb{I}}\label{spinorialconnection}
\end{eqnarray}
called spinorial connection. This is needed to write
\begin{eqnarray}
&\boldsymbol{\nabla}_{\mu}\psi\!=\!\partial_{\mu}\psi
\!+\!\boldsymbol{\Omega}_{\mu}\psi\label{spincovder}
\end{eqnarray}
as spinorial covariant derivative. As well known, the commutator of spinorial covariant derivatives can justify the definitions of space-time and gauge tensors given by 
\begin{eqnarray}
&R^{i}_{\phantom{i}j\mu\nu}\!=\!\partial_{\mu}\Omega^{i}_{\phantom{i}j\nu}
\!-\!\partial_{\nu}\Omega^{i}_{\phantom{i}j\mu}
\!+\!\Omega^{i}_{\phantom{i}k\mu}\Omega^{k}_{\phantom{k}j\nu}
\!-\!\Omega^{i}_{\phantom{i}k\nu}\Omega^{k}_{\phantom{k}j\mu}\\
&F_{\mu\nu}\!=\!\partial_{\mu}A_{\nu}\!-\!\partial_{\nu}A_{\mu}
\end{eqnarray}
that is the Riemann curvature and the Maxwell strength.

When the polar form is taken into account, and considering that we can formally write the expansion
\begin{eqnarray}
&\boldsymbol{S}\partial_{\mu}\boldsymbol{S}^{-1}\!=\!i\partial_{\mu}\alpha\mathbb{I}
\!+\!\frac{1}{2}\partial_{\mu}\theta_{ij}\boldsymbol{\sigma}^{ij}\label{spintrans}
\end{eqnarray}
then with (\ref{spinorialconnection}) we can define
\begin{eqnarray}
&\partial_{\mu}\alpha\!-\!qA_{\mu}\!\equiv\!P_{\mu}\label{P}\\
&\partial_{\mu}\theta_{ij}\!-\!\Omega_{ij\mu}\!\equiv\!R_{ij\mu}\label{R}
\end{eqnarray}
which can be proven to be tensors and invariant under a gauge transformation simultaneously. With them we can write the spinorial covariant derivative in simple form as
\begin{eqnarray}
&\!\!\!\!\!\!\!\!\boldsymbol{\nabla}_{\mu}\psi\!=\!(-\frac{i}{2}\nabla_{\mu}\beta\boldsymbol{\pi}
\!+\!\nabla_{\mu}\ln{\phi}\mathbb{I}
\!-\!iP_{\mu}\mathbb{I}\!-\!\frac{1}{2}R_{ij\mu}\boldsymbol{\sigma}^{ij})\psi
\label{decspinder}
\end{eqnarray}
from which we also have
\begin{eqnarray}
&\nabla_{\mu}s_{i}\!=\!R_{ji\mu}s^{j}\label{ds}\\
&\nabla_{\mu}u_{i}\!=\!R_{ji\mu}u^{j}\label{du}
\end{eqnarray}
are general geometric identities. Taking the commutator of the spinor field, or of the velocity or spin, yields
\begin{eqnarray}
\!\!\!\!&qF_{\mu\nu}\!=\!-(\nabla_{\mu}P_{\nu}\!-\!\nabla_{\nu}P_{\mu})\label{Maxwell}\\
&\!\!\!\!\!\!\!\!R^{i}_{\phantom{i}j\mu\nu}\!=\!-(\nabla_{\mu}R^{i}_{\phantom{i}j\nu}
\!-\!\!\nabla_{\nu}R^{i}_{\phantom{i}j\mu}
\!\!+\!R^{i}_{\phantom{i}k\mu}R^{k}_{\phantom{k}j\nu}
\!-\!R^{i}_{\phantom{i}k\nu}R^{k}_{\phantom{k}j\mu})\label{Riemann}
\end{eqnarray}
in terms of the Riemann curvature and Maxwell strength, so that they encode electrodynamic and gravitational information as usual. As we said above, when we write the spinor field in its polar form, the spinor field is reconfigured so that its degrees of freedom are isolated from the components transferable into gauge and frames through the phase $\alpha$ and the parameters $\theta_{ij}$ in general. When the phase and parameters are added to gauge potential and spin connection, they do not alter their information, and thus (\ref{P}, \ref{R}) contain the same information of the gauge potential and the spin connection themselves, although in the combination all non-covariant features cancel, so that (\ref{P}, \ref{R}) are gauge invariant and Lorentz covariant, and for this reason they have been called gauge-invariant vector momentum and tensorial connection \cite{Fabbri:2018crr}.

This background still has no torsion. The only reason for this is that so far we have been dealing with interactions that arise from gauging some global transformation, while torsion arises from general geometric considerations alone. As such, it can always be split away from the rest of the connection. Therefore, any theory with torsion can equivalently be developed as a theory without torsion but complemented with a very specific form of torsional interaction in the dynamical sector. The Lagrangian is
\begin{eqnarray}
&\mathscr{L}\!=\!i\overline{\psi}\boldsymbol{\gamma}^{\mu}\boldsymbol{\nabla}_{\mu}\psi
\!-\!XW_{\mu}\overline{\psi}\boldsymbol{\gamma}^{\mu}\boldsymbol{\pi}\psi
\!-\!m\overline{\psi}\psi
\end{eqnarray}
where $W_{\mu}$ is the axial-vector dual of the completely antisymmetric torsion and $X$ the most general torsion-spin coupling constant. If we wrote all quantities in torsionful connection we would have obtained the same results obtained by writing all quantities in torsionless connection so long as we kept the torsion-spin interaction in its most general form. The field equations are
\begin{eqnarray}
&i\boldsymbol{\gamma}^{\mu}\boldsymbol{\nabla}_{\mu}\psi
\!-\!XW_{\mu}\boldsymbol{\gamma}^{\mu}\boldsymbol{\pi}\psi\!-\!m\psi\!=\!0
\label{D}
\end{eqnarray}
called Dirac equations. Multiplying (\ref{D}) by $\boldsymbol{\gamma}^{a}$ and $\boldsymbol{\gamma}^{a}\boldsymbol{\pi}$ and by $\overline{\psi}$ and splitting real and imaginary parts gives
\begin{eqnarray}
\nonumber
&i(\overline{\psi}\boldsymbol{\nabla}^{\alpha}\psi
\!-\!\boldsymbol{\nabla}^{\alpha}\overline{\psi}\psi)
\!-\!\nabla_{\mu}M^{\mu\alpha}-\\
&-XW_{\sigma}M_{\mu\nu}\varepsilon^{\mu\nu\sigma\alpha}\!-\!2mU^{\alpha}\!=\!0
\label{vr}\\
\nonumber
&\nabla_{\alpha}\Phi
\!-\!2(\overline{\psi}\boldsymbol{\sigma}_{\mu\alpha}\!\boldsymbol{\nabla}^{\mu}\psi
\!-\!\boldsymbol{\nabla}^{\mu}\overline{\psi}\boldsymbol{\sigma}_{\mu\alpha}\psi)+\\
&+2X\Theta W_{\alpha}\!=\!0\label{vi}
\end{eqnarray}
\begin{eqnarray}
\nonumber
&\nabla_{\nu}\Theta\!-\!
2i(\overline{\psi}\boldsymbol{\sigma}_{\mu\nu}\boldsymbol{\pi}\boldsymbol{\nabla}^{\mu}\psi\!-\!
\boldsymbol{\nabla}^{\mu}\overline{\psi}\boldsymbol{\sigma}_{\mu\nu}\boldsymbol{\pi}\psi)-\\
&-2X\Phi W_{\nu}\!+\!2mS_{\nu}\!=\!0\label{ar}\\
\nonumber
&(\boldsymbol{\nabla}_{\alpha}\overline{\psi}\boldsymbol{\pi}\psi
\!-\!\overline{\psi}\boldsymbol{\pi}\boldsymbol{\nabla}_{\alpha}\psi)
\!-\!\frac{1}{2}\nabla^{\mu}M^{\rho\sigma}\varepsilon_{\rho\sigma\mu\alpha}+\\
&+2XW^{\mu}M_{\mu\alpha}\!=\!0\label{ai}
\end{eqnarray}
which are called Gordon decompositions.

When in (\ref{vi}, \ref{ar}) we plug the polar form we obtain
\begin{eqnarray}
&\!\!\!\!B_{\mu}\!-\!2P^{\iota}u_{[\iota}s_{\mu]}\!+\!(\nabla\beta\!-\!2XW)_{\mu}
\!+\!2s_{\mu}m\cos{\beta}\!=\!0\label{dep1}\\
&\!\!\!\!R_{\mu}\!-\!2P^{\rho}u^{\nu}s^{\alpha}\varepsilon_{\mu\rho\nu\alpha}\!+\!2s_{\mu}m\sin{\beta}
\!+\!\nabla_{\mu}\ln{\phi^{2}}\!=\!0\label{dep2}
\end{eqnarray}
with $R_{\mu a}^{\phantom{\mu a}a}\!=\!R_{\mu}$ and $\frac{1}{2}\varepsilon_{\mu\alpha\nu\iota}R^{\alpha\nu\iota}\!=\!B_{\mu}$ and which could be proven to be equivalent to polar form of Dirac equations as it has been discussed in \cite{Fabbri:2016laz}. The spinor equations (\ref{D}) consist of $4$ complex equations, that is $8$ real equations, which are as many as the $2$ vectorial equations given by the (\ref{dep1}, \ref{dep2}) above, so such vectorial equations specify all space-time derivatives of both degrees of freedom given by module and Yvon-Takabayashi angle. It is also important to notice that the YT angle, in its being the phase difference between chiral projections, must be expected to be related to the mass term, and here it clearly is.

As one final comment, we would focus on the Maxwell strength and Riemann curvature, which are well known to encode the electrodynamic and gravitational information as a whole. As such, they act as some sort of filters that keep out all information due to gauge and frames.

On the other hand, (\ref{P}, \ref{R}) do have information about electrodynamics and gravity as well as gauge and frames while still being fully covariant. In fact, it is possible to prove that for the conditions that are given by
\begin{eqnarray}
&\nabla_{\mu}P_{\nu}\!-\!\nabla_{\nu}P_{\mu}\!=\!0\\
&\nabla_{\mu}R^{i}_{\phantom{i}j\nu}\!-\!\!\nabla_{\nu}R^{i}_{\phantom{i}j\mu}
\!\!+\!R^{i}_{\phantom{i}k\mu}R^{k}_{\phantom{k}j\nu}
\!-\!R^{i}_{\phantom{i}k\nu}R^{k}_{\phantom{k}j\mu}\!=\!0
\end{eqnarray}
there exist solutions that are non-zero. What this means is that it is possible to find situations that have no gravity nor electrodynamics but for which the information linked to frames and gauge is non-trivial although covariant \cite{Fabbri:2019kfr}.

General comments about the polar decomposition of the spinorial fields can also be found in reference \cite{Fabbri:2017pwp}.

A summary of previous results is in reference \cite{Fabbri:2020ypd}.

Quantum systems have been investigated in \cite{Fabbri:2019tad}.
%%%%%%%%%%%%%%%%%%%%%%%%%%%%%%%%%%%%%%%%%%%%%%%%%%%%%%%%%%%%%%%%%%%%%%%%%%%%%%%%%%%%%%%%%%%%%%%%%%%
%%%%%%%%%%%%%%%%%%%%%%%%%%%%%%%%%%%%%%%%%%%%%%%%%%%%%%%%%%%%%%%%%%%%%%%%%%%%%%%%%%%%%%%%%%%%%%%%%%%
\section{Re-normalizable Chern-Simons Propagating Torsion}
Having presented the general theory of spinor fields in polar form, we can now work out the details to construct a re-normalizable Chern-Simons extension of the propagating torsion theory. To this purpose we need to assign the dynamics of torsion, straightforwardly given by
\begin{eqnarray}
&\mathscr{L}\!=\!-\frac{1}{4}(\partial W)_{\mu\nu}(\partial W)^{\mu\nu}
\!+\!\frac{1}{2}M^{2}W^{\mu}W_{\mu}
\end{eqnarray}
with $(\partial W)_{\mu\nu}$ curl of the axial-vector torsion and $M$ mass of torsion, and this Lagrangian is re-normalizable. Since we want torsion coupled to spinors, we must take the full
\begin{eqnarray}
\nonumber
&\mathscr{L}\!=\!i\overline{\psi}\boldsymbol{\gamma}^{\mu}\boldsymbol{\nabla}_{\mu}\psi
-\frac{1}{4}(\partial W)_{\mu\nu}(\partial W)^{\mu\nu}-\\
&-XW_{\mu}\overline{\psi}\boldsymbol{\gamma}^{\mu}\boldsymbol{\pi}\psi
\!+\!\frac{1}{2}M^{2}W^{\mu}W_{\mu}\!-\!m\overline{\psi}\psi
\end{eqnarray}
or in polar form
\begin{eqnarray}
\nonumber
&\mathscr{L}\!=\!2\phi^{2}[\frac{1}{2}s^{\mu}
(\nabla_{\mu}\beta\!-\!2XW_{\mu}\!+\!B_{\mu})\!+\!u^{\mu}P_{\mu}\!-\!m\cos{\beta}]-\\
&-\frac{1}{4}(\partial W)_{\mu\nu}(\partial W)^{\mu\nu}\!+\!\frac{1}{2}M^{2}W^{2}
\end{eqnarray}
which is still re-normalizable \cite{Fabbri:2014dxa}. In fact, it is the Proca Lagrangian for an axial-vector field coupled to spinors.

To include all re-normalizable Chern-Simons terms, we have to add a $4$ mass dimensional term of topological-like structure, that is of the form $b\nabla_{\mu}K^{\mu}$ with $b$ pseudo-scalar and $K^{\mu}$ axial-vector. The most straightforward term is of course the Chern-Simons contribution given by
\begin{eqnarray}
&\mathscr{L}\!=\!\frac{A}{2}b
\nabla_{\mu}[W_{\nu}(\partial W)_{\rho\sigma}\varepsilon^{\mu\nu\rho\sigma}]
\end{eqnarray}
as it is clear. A term with less derivatives but of the same mass dimension can be included in terms of
\begin{eqnarray}
&\mathscr{L}\!=\!\frac{B}{2}b\nabla_{\mu}(W^{\mu}W^{2})
\end{eqnarray}
and as a quick inventory would show nothing more can be added at this mass dimension. To lower mass dimension we can still add the term that is given by
\begin{eqnarray}
&\mathscr{L}\!=\!kb\nabla_{\mu}W^{\mu}
\end{eqnarray}
itself. Altogether, they can be written according to
\begin{eqnarray}
&\!\!\!\!\mathscr{L}\!=\!b\nabla_{\mu}[\frac{A}{2}W_{\nu}(\partial W)_{\rho\sigma}
\varepsilon^{\mu\nu\rho\sigma}\!+\!\frac{B}{2}W^{\mu}W^{2}\!+\!kW^{\mu}]
\end{eqnarray}
having factorized the $b$ out. The unitary mass dimension of torsion implies that the divergence of the axial-vector is already $4$ mass dimensional, so re-normalizability must require a pseudo-scalar with zero mass dimension, which now we know to exist, and it is simply given by the aforementioned Yvon-Takabayashi angle. Therefore we have
\begin{eqnarray}
&\!\!\!\!\mathscr{L}\!=\!-\beta\nabla_{\mu}[\frac{A}{2}W_{\nu}(\partial W)_{\rho\sigma}
\varepsilon^{\mu\nu\rho\sigma}\!+\!\frac{B}{2}W^{\mu}W^{2}\!+\!kW^{\mu}]
\end{eqnarray}
as the most comprehensive Chern-Simons term compatible with re-normalizability. And eventually we have that
\begin{eqnarray}
\nonumber
&\mathscr{L}\!=\!2\phi^{2}[\frac{1}{2}s^{\mu}
(\nabla_{\mu}\beta\!-\!2XW_{\mu}\!+\!B_{\mu})\!+\!u^{\mu}P_{\mu}\!-\!m\cos{\beta}]-\\
\nonumber
&-\beta\nabla_{\mu}[\frac{A}{2}W_{\nu}(\partial W)_{\rho\sigma}\varepsilon^{\mu\nu\rho\sigma}
\!+\!\frac{B}{2}W^{\mu}W^{2}\!+\!kW^{\mu}]-\\
&-\frac{1}{4}(\partial W)_{\mu\nu}(\partial W)^{\mu\nu}\!+\!\frac{1}{2}M^{2}W^{2}
\end{eqnarray}
is the most general re-normalizable Chern-Simons extension of the propagating torsion field Lagrangian.

Variation of this Lagrangian gives the following
\begin{eqnarray}
\nonumber
&\nabla_{\mu}(\partial W)^{\mu\nu}\!+\!M^{2}W^{\nu}\!=\!
-\nabla_{\alpha}\beta[A(\partial W)_{\rho\sigma}\varepsilon^{\rho\sigma\alpha\nu}+\\
&+B(W^{\alpha}W^{\nu}\!+\!\frac{1}{2}g^{\alpha\nu}W^{2})\!+\!kg^{\alpha\nu}]\!+\!2X\phi^{2}s^{\nu}
\label{p}
\end{eqnarray}
and
\begin{eqnarray}
&-2P^{\iota}u_{[\iota}s_{\mu]}\!+\!B_{\mu}\!-\!2XW_{\mu}
\!+\!\nabla_{\mu}\beta\!+\!2s_{\mu}m\cos{\beta}\!=\!0\label{1}\\
\nonumber
&-s_{\mu}\nabla_{\alpha}[\frac{A}{2}W_{\nu}(\partial W)_{\rho\sigma}
\varepsilon^{\rho\sigma\alpha\nu}
\!+\!\frac{B}{2}W^{\alpha}W^{2}\!+\!kW^{\alpha}]/\phi^{2}+\\
&+\nabla_{\mu}\ln{\phi^{2}}\!+\!R_{\mu}
\!-\!2P^{\rho}u^{\nu}s^{\alpha}\varepsilon_{\mu\rho\nu\alpha}
\!+\!2s_{\mu}m\sin{\beta}\!=\!0\label{2}
\end{eqnarray}
as the set of field equations.

By taking the divergence of (\ref{p}) one gets
\begin{eqnarray}
\nonumber
&\!\!\!\!(M^{2}\!+\!BW\!\!\cdot\!\nabla\beta)\nabla\!\!\cdot\!W\!\!=\!
-BW_{\nu}\nabla_{\alpha}\beta(\nabla^{\nu}W^{\alpha}\!\!+\!\!\nabla^{\alpha}W^{\nu})-\\
&-\nabla_{\alpha}\nabla_{\nu}\beta[B(W^{\alpha}W^{\nu}\!+\!\frac{1}{2}g^{\alpha\nu}W^{2})
\!+\!kg^{\alpha\nu}]\!+\!\!X\nabla\!\!\cdot\!S
\label{a}
\end{eqnarray}
and by contracting (\ref{2}) with $s_{\mu}$ one gets
\begin{eqnarray}
\nonumber
&\!\!\nabla\!\cdot\!S\!=\!4\phi^{2}m\sin{\beta}
\!-\!\frac{A}{2}(\partial W)_{\mu\nu}(\partial W)_{\rho\sigma}\varepsilon^{\mu\nu\rho\sigma}-\\
&-BW_{\mu}W_{\nu}(\nabla^{\nu}W^{\mu}\!+\!\nabla^{\mu}W^{\nu})\!-\!(BW^{2}\!+\!2k)\nabla\!\cdot\!W
\label{b}
\end{eqnarray}
as the partially conserved axial-vector current.
%%%%%%%%%%%%%%%%%%%%%%%%%%%%%%%%%%%%%%%%%%%%%%%%%%%%%%%%%%%%%%%%%%%%%%%%%%%%%%%%%%%%%%%%%%%%%%%%%%%
%%%%%%%%%%%%%%%%%%%%%%%%%%%%%%%%%%%%%%%%%%%%%%%%%%%%%%%%%%%%%%%%%%%%%%%%%%%%%%%%%%%%%%%%%%%%%%%%%%%
\section{Causal Structure Analysis}
In the previous section, we obtained the field equations of the re-normalizable CS extension for the propagating torsion theory. Nevertheless, these field equations contain non-linear torsion terms that, quite generally, might interfere with the propagation of torsion, as described by the Velo-Zwanziger analysis \cite{Velo:1970ur}. It is therefore necessary to perform such an analysis to make sure that this does not happen, or to remove all terms spoiling causality.

Mathematically, our goal is to invert (\ref{a}, \ref{b}) to extract $\nabla\!\cdot\!W$ which will have to be substituted into (\ref{p}) to give the field equations in the form that encodes its primary constraint. Dropping every contribution but the highest-order derivative terms, we get the characteristic equation
\begin{eqnarray}
\nonumber
&[(M^{2}\!+\!2Xk\!+\!BW\!\!\cdot\!\nabla\beta\!+\!BXW^{2})n^{2}g^{\sigma\nu}+\\
\nonumber
&+B(n\!\cdot\!W\nabla^{\nu}\beta\!+\!n\!\cdot\!\nabla\beta W^{\nu}+\\
\nonumber
&+2Xn\!\cdot\!WW^{\nu})n^{\sigma}+\\
&+2AX(\partial W)_{\rho\beta}\varepsilon^{\rho\beta\mu\nu}n_{\mu}n^{\sigma}]W_{\nu}\!=\!0
\end{eqnarray}
which is to be valid for any $W^{\nu}$ solution of the field equations. Then the characteristic determinant is
\begin{eqnarray}
\nonumber
&\mathrm{det}|[M^{2}\!+\!2Xk\!+\!BW\!\cdot\!(\nabla\beta\!+\!XW)]n^{2}g^{\sigma\nu}+\\
\nonumber
&+[BW\!\!\cdot\!n(\nabla\beta\!+\!XW)^{\nu}+\\
\nonumber
&+B(\nabla\beta\!+\!XW)\!\cdot\!n W^{\nu}+\\
&+2AX(\partial W)_{\rho\beta}\varepsilon^{\rho\beta\mu\nu}n_{\mu}]n^{\sigma}|\!=\!0
\end{eqnarray}
that is with form $\mathrm{det}|E^{\nu\sigma}|\!=\!\mathrm{det}|g^{\nu\sigma}\!+\!F^{\nu}G^{\sigma}|\!=\!0$ as clear.

The computation of this determinant can be done by simply looking for an inverse. We will look for an inverse in the form $(E^{-1})^{\mu\nu}\!=\!g^{\mu\nu}\!-\!KF^{\mu}G^{\nu}$ and it is easy to see that $K(1\!+\!G\!\cdot\!F)\!=\!1$ must hold. Then the inverse is given by $(E^{-1})^{\mu\nu}\!=\!g^{\mu\nu}\!-\!(1\!+G\!\cdot\!F)^{-1}F^{\mu}G^{\nu}$ and the determinant is $\mathrm{det}|E^{\mu \nu}|\!=\!(1\!+G\!\cdot\!F)$ straightforwardly. We can hence compute the characteristic determinant as given by
\begin{eqnarray}
\nonumber
&[M^{2}\!+\!2Xk\!+\!BW\!\cdot\!(\nabla\beta\!+\!XW)]n^{2}+\\
&+2BW\!\!\cdot\!n(\nabla\beta\!+\!XW)\!\cdot\!n\!=\!0
\end{eqnarray}
as it can be checked with a direct substitution.

The causal structure is therefore always determined in terms of the solutions themselves, and for a weak torsion we may approximate the above with
\begin{eqnarray}
&(M^{2}\!+\!2Xk)n^{2}\!+\!2BW\!\!\cdot\!n(\nabla\beta\!+\!XW)\!\cdot\!n\!=\!0
\end{eqnarray}
showing that $n^{2}\!>\!0$ is always possible. As a consequence, wave fronts may always escape the light-cone constraint.

The light-cone structure is however always respected if
\begin{eqnarray}
&BW\!\!\cdot\!n(\nabla\beta\!+\!XW)\!\cdot\!n\!=\!0
\end{eqnarray}
which is valid for any solution $W_{\alpha}$ only in the case where we have that $B\!=\!0$ identically as a restriction.
%%%%%%%%%%%%%%%%%%%%%%%%%%%%%%%%%%%%%%%%%%%%%%%%%%%%%%%%%%%%%%%%%%%%%%%%%%%%%%%%%%%%%%%%%%%%%%%%%%%
%%%%%%%%%%%%%%%%%%%%%%%%%%%%%%%%%%%%%%%%%%%%%%%%%%%%%%%%%%%%%%%%%%%%%%%%%%%%%%%%%%%%%%%%%%%%%%%%%%%
\section{Field Equations}
Having performed such a reduction, we have that the most general re-normalizable Chern-Simons extension of the causally propagating torsion field is given by
\begin{eqnarray}
\nonumber
&\mathscr{L}\!=\!2\phi^{2}[\frac{1}{2}s^{\mu}
(\nabla_{\mu}\beta\!-\!2XW_{\mu}\!+\!B_{\mu})\!+\!u^{\mu}P_{\mu}\!-\!m\cos{\beta}]-\\
\nonumber
&-\beta[\frac{A}{4}(\partial W)_{\mu\nu}(\partial W)_{\rho\sigma}\varepsilon^{\mu\nu\rho\sigma}
\!+\!k\nabla_{\mu}W^{\mu}]-\\
&-\frac{1}{4}(\partial W)_{\mu\nu}(\partial W)^{\mu\nu}\!+\!\frac{1}{2}M^{2}W^{2}
\end{eqnarray}
where we have used the polar form for the spinor sector.

The corresponding system of field equations is
\begin{eqnarray}
\nonumber
&\nabla_{\mu}(\partial W)^{\mu\nu}\!+\!M^{2}W^{\nu}=\\
&=-\nabla_{\alpha}\beta[A(\partial W)_{\rho\sigma}\varepsilon^{\rho\sigma\alpha\nu}
\!+\!kg^{\alpha\nu}]\!+\!2X\phi^{2}s^{\nu}
\end{eqnarray}
and
\begin{eqnarray}
&-2P^{\iota}u_{[\iota}s_{\mu]}\!+\!B_{\mu}\!-\!2XW_{\mu}
\!+\!\nabla_{\mu}\beta\!+\!2s_{\mu}m\cos{\beta}\!=\!0\\
\nonumber
&-s_{\mu}[\frac{A}{4}(\partial W)_{\alpha\nu}(\partial W)_{\rho\sigma}
\varepsilon^{\rho\sigma\alpha\nu}\!+\!k\nabla_{\alpha}W^{\alpha}]/\phi^{2}+\\
&+\nabla_{\mu}\ln{\phi^{2}}\!+\!R_{\mu}
\!-\!2P^{\rho}u^{\nu}s^{\alpha}\varepsilon_{\mu\rho\nu\alpha}
\!+\!2s_{\mu}m\sin{\beta}\!=\!0
\end{eqnarray}
in which all the torsion non-linearities have been lost.

They develop the divergences
\begin{eqnarray}
&M^{2}\nabla\!\cdot\!W\!=-k\nabla^{2}\beta\!+\!\!X\nabla\!\!\cdot\!S
\end{eqnarray}
and
\begin{eqnarray}
\nonumber
&\nabla S\!=\!4\phi^{2}m\sin{\beta}-\\
&-\frac{A}{2}(\partial W)_{\mu\nu}(\partial W)_{\rho\sigma}\varepsilon^{\mu\nu\rho\sigma}
\!-\!2k\nabla\!\cdot\!W
\end{eqnarray}
where the mixed terms are now easier to manage.

Next step would be finding solutions, but because this is extremely complicated we will try to set some assumption and approximation. A first thing to notice is that in absence of electrodynamics the momentum $P_{\nu}$ consists in a pure gauge, and since for spinors in eigen-state of spin around the third axis phases are equivalent to rotations around the third axis then all information in $P_{\nu}$ could be described in $R_{12\nu}$ instead \cite{Fabbri:2020ypd,Fabbri:2019tad}. Hence there is no loss of generality in setting $P_{\nu}\!=\!0$ identically. And as is usual for massive fields like torsion, we might take the effective approximation. In this instance we have to consider that all the torsional kinetic terms are negligible compared to any other term, so the above equations reduce to
\begin{eqnarray}
&M^{2}W_{\mu}\!=\!-k\nabla_{\mu}\beta\!+\!2X\phi^{2}s_{\mu}
\end{eqnarray}
and
\begin{eqnarray}
&B_{\mu}\!-\!2XW_{\mu}\!+\!\nabla_{\mu}\beta\!+\!2s_{\mu}m\cos{\beta}\!=\!0\\
&-s_{\mu}k\nabla_{\alpha}W^{\alpha}/\phi^{2}\!+\!\nabla_{\mu}\ln{\phi^{2}}
\!+\!R_{\mu}\!+\!2s_{\mu}m\sin{\beta}\!=\!0
\end{eqnarray}
with the equations for the torsion axial-vector being those found in \cite{Mercuri:2009zi} and \cite{Lattanzi:2009mg}, and later discussed also in \cite{Castillo-Felisola:2015ema}.

The divergence of the torsion axial-vector remains as
\begin{eqnarray}
&M^{2}\nabla\!\cdot\!W\!=-k\nabla^{2}\beta\!+\!\!X\nabla\!\!\cdot\!S
\end{eqnarray}
but now the divergence of the spin is
\begin{eqnarray}
&\nabla S\!=\!4\phi^{2}m\sin{\beta}\!-\!2k\nabla\!\cdot\!W
\end{eqnarray}
simplifying the partially conserved axial-vector current.

In effective approximation, it is now possible to substitute torsion everywhere remaining with
\begin{eqnarray}
\nonumber
&-s_{\mu}(4X^{2}M^{-2}\phi^{2}\!-\!2m\cos{\beta})+\\
&+B_{\mu}\!+\!(1\!+\!2kXM^{-2})\nabla_{\mu}\beta\!=\!0\\
\nonumber
&s_{\mu}(2M^{2}m\sin{\beta}\!+\!k^{2}\nabla^{2}\beta/\phi^{2})+\\
&+(M^{2}+2Xk)(R_{\mu}\!+\!\nabla_{\mu}\ln{\phi^{2}})\!=\!0
\end{eqnarray}
where only spinor degrees of freedom are present.

It is not possible to perform any more approximations on torsion, but we can still assume the Yvon-Takabayashi angle to be small and see what happens. In this case we can have the field equation for the YT angle plugged into the field equation for the module and set $\beta\!\rightarrow\!0$ getting
\begin{eqnarray}
\nonumber
&(M^{2}\!+\!2Xk)^{2}(\phi^{2}R\!+\!\nabla\phi^{2})_{\mu}+\\
\nonumber
&+4X^{2}k^{2}(\phi^{2}R\!+\!\nabla\phi^{2})\!\cdot\!s s_{\mu}-\\
&-s_{\mu}M^{2}k^{2}(\nabla B\!+\!2msR)\!=\!0
\end{eqnarray}
which is in terms of the module alone.

This equation can be projected along the main directions consequently decomposing into
\begin{eqnarray}
&(M^{2}\!+\!4Xk)(\phi^{2}R\!+\!\nabla\phi^{2})\!\cdot\!s\!+\!k^{2}(\nabla B\!+\!2msR)\!=\!0
\end{eqnarray}
\begin{eqnarray}
&Ru\phi^{2}\!+\!u^{\mu}\nabla_{\mu}\phi^{2}\!=\!0
\end{eqnarray}
all of which being irreducible.

Notice that vanishing module is not a solution and one minimum for $\phi^{2}$ is reached whenever it is
\begin{eqnarray}
\langle\phi^{2}/m\rangle
\!=\!-\frac{2k^{2}}{M^{2}\!+\!4Xk}\left(1\!+\!\frac{\nabla\!\cdot\!B}{2msR}\right)
\end{eqnarray}
in terms of the constants and the tensorial connection.

Notice that even in the case in which the tensorial connection were to vanish we would still remain with
\begin{eqnarray}
\langle\phi^{2}/m\rangle\!=\!-\frac{2k^{2}}{M^{2}\!+\!4Xk}
\end{eqnarray}
in terms of topological parameter and torsion constants.

The persistence of a non-trivial vacuum for the spinor field even at infinity is to be tied to the correspondingly non-trivial topological features displayed at infinity.
%%%%%%%%%%%%%%%%%%%%%%%%%%%%%%%%%%%%%%%%%%%%%%%%%%%%%%%%%%%%%%%%%%%%%%%%%%%%%%%%%%%%%%%%%%%%%%%%%%%
%%%%%%%%%%%%%%%%%%%%%%%%%%%%%%%%%%%%%%%%%%%%%%%%%%%%%%%%%%%%%%%%%%%%%%%%%%%%%%%%%%%%%%%%%%%%%%%%%%%
\section{Conclusion}
In this paper, we have discussed in what way it is possible to exploit the polar form of spinor fields to construct a re-normalizable Chern-Simons extension for the propagating torsion theory, and we have built the most general case that respects the causal structure. We obtained that in such a case the Lagrangian has a term of type $\beta\nabla_{\mu}K^{\mu}$ with $K^{\mu}$ having two contributions, one being the torsion $W^{\mu}$ and one being $W_{\nu}(\partial W)_{\rho\sigma}\varepsilon^{\mu\nu\rho\sigma}$ which possesses the form of the spin density tensor of the torsion field. Since torsion is sourced by the spin density then the spin density of torsion may give rise to torsional self-interactions that are not present in the torsion field equations, though they are present in this topological extension. This fact could be interpreted quite easily if we think that the spin might well be in its origin a purely topological concept.

We have then investigated the case in which the torsion tensor is taken in the effective approximation, that is the limit in which the torsion axial-vector loses all dynamical contributions against the lowest-mass dimensional terms given by the mass term and the topological term proportional to the $k$ constant. By integrating out the torsion axial-vector first, and the Yvon-Takabayashi angle later, we have actually been capable of isolating the field equation for the module, displaying the singular character of possessing no solution that would vanish at the infinity.

We regard this as one of the features that indicates in a clear way how the theory is sensitive to the topological features displayed at the boundary of the space-time.

Further investigations should concern the possibility to study the vacuum of the torsion tensor.
%%%%%%%%%%%%%%%%%%%%%%%%%%%%%%%%%%%%%%%%%%%%%%%%%%%%%%%%%%%%%%%%%%%%%%%%%%%%%%%%%%%%%%%%%%%%%%%%%%%
%%%%%%%%%%%%%%%%%%%%%%%%%%%%%%%%%%%%%%%%%%%%%%%%%%%%%%%%%%%%%%%%%%%%%%%%%%%%%%%%%%%%%%%%%%%%%%%%%%%

%%%%%%%%%%%%%%%%%%%%%%%%%%%%%%%%%%%%%%%%%%%%%%%%%%%%%%%%%%%%%%%%%%%%%%%%%%%%%%%%%%%%%%%%%%%%%%%%%%%

\begin{thebibliography}{40}
\bibitem{Hehl:1976kj}
F.W.Hehl, P.Von Der Heyde, G.D.Kerlick, J.M.Nester,\\
``General Relativity with Spin and Torsion: Foundations\\
and Prospects'', \textit{Rev.Mod.Phys.}\textbf{48}, 393 (1976).
\bibitem{Shapiro:2001rz}
I.L.Shapiro, ``Physical aspects of the space-time torsion'',\\
\textit{Phys.Rept.}\textbf{357}, 113 (2002).
\bibitem{Hammond:2002rm}
R.T.Hammond, ``Torsion gravity'',\\
\textit{Rept.Prog.Phys.}\textbf{65}, 599 (2002).
\bibitem{Arcos:2005ec}
H.I.Arcos, J.G.Pereira, ``Torsion gravity:\\ A Reappraisal'', \textit{Int.J.Mod.Phys.D}\textbf{13}, 2193 (2004).
\bibitem{Laemmerzahl:1993zn}
C.Laemmerzahl, A.Macias, ``On the dimensionality of\\ space-time'', \textit{J. Math. Phys.} \textbf{34}, 4540 (1993).
\bibitem{Audretsch:1988tu}
J.Audretsch, C.Lammerzahl, ``Constructive\\ Axiomatic Approach To Space-time Torsion'',\\ \textit{Class. Quant. Grav.} \textbf{5}, 1285 (1988).
\bibitem{Fabbri:2006xq} 
L.Fabbri, ``On a completely antisymmetric Cartan torsion\\ tensor'', \textit{In Annales de la Fondation de Broglie,\\ Special Issue on Torsion (2007)}.
\bibitem{Fabbri:2009se}
L.Fabbri, ``On the Principle of Equivalence'',\\ \textit{In Contemporary Fundamental Physics,\\ Einstein and Hilbert: Dark Matter (2012)}
\bibitem{Fabbri:2008rq} 
L.Fabbri, ``On the problem of Unicity in Einstein-Sciama-Kibble Theory'', \textit{Annales Fond. Broglie}\textbf{33}, 365 (2008).
\bibitem{Fabbri:2009yc} 
L.Fabbri, ``On the consistency of Constraints in Matter Field Theories'', \textit{Int.J.Theor.Phys.}\textbf{51}, 954 (2012).
\bibitem{Fabbri:2014naa} 
L.Fabbri, ``Least-order torsion-gravity for fermion fields,\\ and the nonlinear potentials in the standard models'',\\ \textit{Int.J.Geom.Meth.Mod.Phys.}\textbf{11}, 1450073 (2014).
\bibitem{Fabbri:2017rjf}
L.Fabbri, ``Singularity-free spinors in gravity with propagating torsion'',
\textit{Mod.Phys.Lett.A}\textbf{32}, 1750221 (2017).
\bibitem{Fabbri:2017xch}
L.Fabbri, ``A geometrical assessment of spinorial energy conditions'', \textit{Eur.Phys.J.Plus}\textbf{132}, 156 (2017).
\bibitem{Fabbri:2010rw}
L.Fabbri, ``On geometric relativistic foundations of\\ matter field equations and plane wave solutions'', \\ \textit{Mod.Phys.Lett.A}\textbf{27}, 1250028 (2012).
\bibitem{Fabbri:2012ag}
L.Fabbri, ``On a purely geometric approach to the\\ Dirac matter field and its quantum properties'',\\ \textit{Int.J.Theor.Phys.}\textbf{53}, 1896 (2014).
\bibitem{Fabbri:2014dxa}
L.Fabbri, ``A discussion on the most general torsion-gravity with electrodynamics for Dirac spinor matter\\ fields'', \textit{Int.J.Geom.Meth.Mod.Phys.}\textbf{12}, 1550099 (2015).
\bibitem{L}
P.Lounesto, \textit{Clifford Algebras and\\ Spinors} (Cambridge University Press, 2001).
\bibitem{Cavalcanti:2014wia}
R.T.Cavalcanti, ``Classification of Singular Spinor\\ Fields and Other Mass Dimension One Fermions'',\\ \textit{Int.J.Mod.Phys.D}\textbf{23}, 1444002 (2014).
\bibitem{HoffdaSilva:2017waf}
J.M.Hoff da Silva, R.T.Cavalcanti, ``Revealing how\\ different spinors can be: the Lounesto 
spinor\\ classification'', \textit{Mod.Phys.Lett.A}\textbf{32}, 1730032 (2017).
\bibitem{daSilva:2012wp}
J.M.Hoff da Silva, R.da Rocha, ``Unfolding Physics\\ from the Algebraic Classification of Spinor\\ Fields'', \textit{Phys. Lett. B}\textbf{718}, 1519 (2013).
\bibitem{Ablamowicz:2014rpa}
R.Ab{\l}amowicz, I.Gon{\c c}alves, R.da Rocha, ``Bilinear\\ Covariants and Spinor Fields Duality in Quantum\\ Clifford Algebras'', \textit{J. Math. Phys.}\textbf{55}, 103501 (2014).
\bibitem{Rodrigues:2005yz}
W.A.Rodrigues, R.da Rocha, J.Vaz, ``Hidden\\ consequence of active local Lorentz invariance'',\\ \textit{Int.J.Geom.Meth.Mod.Phys.}\textbf{2}, 305 (2005).
\bibitem{HoffdaSilva:2009is} 
J.M.Hoff da Silva, R.da Rocha,
``From Dirac Action to\\ ELKO Action'',
\textit{Int.J.Mod.Phys.A}\textbf{24}, 3227 (2009).
\bibitem{daRocha:2008we}
R.da Rocha, J.M.Hoff da Silva, ``ELKO, flagpole and\\ flag-dipole spinor fields, and the instanton Hopf\\ fibration'', \textit{Adv. Appl. Clifford Algebras} \textbf{20}, 847 (2010).
\bibitem{daRocha:2013qhu}
R.da Rocha,L.Fabbri,J.M.Hoff da Silva,R.T.Cavalcanti, J.A.Silva-Neto,
``Flag-Dipole Spinor Fields in ESK Gravities'',
\textit{J.Math.Phys.}\textbf{54},102505(2013).
\bibitem{Fabbri:2016msm}
L.Fabbri,
``A generally-relativistic gauge\\ classification of the Dirac fields'',\\ \textit{Int.J.Geom.Meth.Mod.Phys.}\textbf{13},1650078(2016).
\bibitem{Fabbri:2018crr} 
L.Fabbri, ``Covariant inertial forces for spinors'',\\ 
\textit{Eur.Phys.J.C}\textbf{78}, 783 (2018).
\bibitem{Fabbri:2016laz}
L.Fabbri,
``Torsion Gravity for Dirac Fields'',\\ 
\textit{Int.J.Geom.Meth.Mod.Phys.}\textbf{14},1750037(2017).
\bibitem{Fabbri:2019kfr} 
L.Fabbri, ``Polar solutions with tensorial connection of\\ the spinor equation'', \textit{Eur.Phys.J.C}\textbf{79}, 188 (2019).
\bibitem{Fabbri:2017pwp}
L.Fabbri, ``General Dynamics of Spinors'',\\ 
\textit{Adv. Appl. Clifford Algebras}\textbf{27}, 2901 (2017).
\bibitem{Fabbri:2020ypd}
L.Fabbri, ``Spinors in Polar Form'', arXiv:2003.10825
\bibitem{Fabbri:2019tad} 
L.Fabbri, ``Geometry, Zitterbewegung, Quantization'',\\ \textit{Int.J.Geom.Meth.Mod.Phys.}\textbf{16}, 1950146 (2019).
\bibitem{Velo:1970ur} 
G.Velo, D.Zwanziger, ``Noncausality and other defects\\ of interaction lagrangians for particles with spin\\ one and higher'', \textit{Phys.Rev.}\textbf{188}, 2218 (1969).
\bibitem{Mercuri:2009zi} 
S.Mercuri, ``Peccei-Quinn mechanism in gravity\\
and the nature of the Barbero-Immirzi parameter'',\\ 
\textit{Phys.Rev.Lett.}\textbf{103}, 081302 (2009).
\bibitem{Lattanzi:2009mg} 
M.Lattanzi, S.Mercuri, ``A solution of the strong CP\\
problem via the Peccei-Quinn mechanism through the\\
Nieh-Yan modified gravity and cosmological\\
implications'', \textit{Phys.Rev.D}\textbf{81},125015 (2010).
\bibitem{Castillo-Felisola:2015ema}
O.Castillo-Felisola, C.Corral, S.Kovalenko, I.Schmidt,\\ 
V.E.Lyubovitskij, ``Axions in gravity with torsion'',\\
\textit{Phys.Rev.D}\textbf{91}, 085017 (2015).
\end{thebibliography}
\end{document}